\documentclass[aps,prl,twocolumn,showpacs,superscriptaddress,floatfix]{revtex4}  
\usepackage[utf8]{inputenc}
\usepackage[T1]{fontenc}
\usepackage{mathptmx}
\usepackage{etoolbox}
\usepackage{graphicx}
\usepackage{dcolumn}
\usepackage{bm}
\usepackage{epstopdf,epsfig}
\usepackage{amssymb}
\usepackage{amsmath}
\usepackage{newtxtext}
\usepackage{newtxmath}
\usepackage{natbib}
\usepackage{hyperref}
\usepackage{xcolor}
\usepackage{adjustbox}
\hypersetup{
    colorlinks = true,
    urlcolor   = blue,
    citecolor  = black,
}

\newcommand{\RomanNumeralCaps}[1]
\linenumbers

\newcommand\Rey{{\rm Re}}
\newcommand\Rm{{\textit{Rm}}}
\newcommand\R{{\textit{R}}}
\newcommand\Pm{{\textit{Pm}}}
\newcommand\Ha{{\textit{Ha}}}

\begin{document}

\title[Inflection point instability in Hartmann channel]{Inflection point instability in Hartmann channel flow with variable electric conductivity}
\author{R. Okatev}
\email{okatev@icmm.ru}
\affiliation{Institute of Continuous Media Mechanics, Korolyov str.1, Perm, 614018, Russia}

\author{O. Zikanov}
\affiliation{University of Michigan - Dearborn, Dearborn, MI 48128-1491, USA}

\author{D. Krasnov}
\affiliation{Technische Universitat Ilmenau, PF 100565, 98684 Ilmenau, Germany}

\author{P. Frick}
\affiliation{Institute of Continuous Media Mechanics, Korolyov str.1, Perm, 614018, Russia}

\begin{abstract}
The stability of a flow of an electrically conducting, incompressible fluid in a channel with an imposed uniform wall-normal magnetic field and electrically insulating walls is studied using linear stability analysis and direct numerical simulations. The novelty of the system, which differentiates it from the classical Hartmann channel flow, is that, as in some technological applications of liquid metals, the electric conductivity and viscosity of the fluid vary across the channel. This variation is found to have a strong influence on the stability characteristics of the flow. Specifically, a linear variation in electric conductivity significantly alters the base velocity profile, leading to pronounced asymmetry and the development of inflection points.  When this transformation is sufficiently strong, the flow becomes linearly unstable at Reynolds numbers much lower than the threshold for the linear instability of the Hartmann channel flow. The instability exhibits distinct features: a large typical axial wavelength and the localization of perturbation growth in the channel core. The characteristics of this instability suggest a mechanism similar to the classical inviscid inflection-point instability of one-dimensional velocity profiles. The resulting transition to turbulence is demonstrated in direct numerical simulations.
\end{abstract}

\maketitle

\section{Introduction}
\label{sec:intro}

In this paper, we revisit the question of the instability of a wall-bounded parallel shear flow of an electrically conducting fluid affected by an imposed steady magnetic field. The novelty of the work is that the physical properties of the fluid, such as its electric conductivity or viscosity, are not required to be constant. In addition to purely theoretical interest in how the variation of physical properties affects instability, the study is motivated by technological applications of liquid metals, such as the development of MHD (magnetohydrodynamic) stirring systems \citep{2014-Mahyd-Denisov-Stirrer,2013-EPhys-Stiller-Stirrer} and MHD separation of impurities. \citep{2013-Mahyd-Kolesnichenko-Separation,2014-MMTB-Zhang-SeparationReview} We can also mention the breeding blankets of nuclear fusion reactors (see, e.g., Refs. \onlinecite{smolentsev:2021} and \onlinecite{mistrangelo:2021}), in particular, the molten salt (e.g., FLiBe) designs for which the strength of the magnetic field effect is comparable to that studied in this paper. 
In all of these flows, the properties of the fluid may vary because of strong gradients of temperature or impurity concentration. Typically, there is a technological need for either the intensification or suppression of mixing. A better understanding and quantitative assessment of the instability and transition to turbulence is, therefore, desirable.

Technological and laboratory MHD flows of liquid metals are typically characterized by small values of the magnetic Reynolds number $\Rm=Uh/\eta$ and the magnetic Prandtl number $\Pm=\nu/\eta$. Here $U$ and  $h$ are the typical velocity and length scales (the mean velocity and half-width of the channel in our case); $\eta=(\sigma \mu_0)^{-1}$ is the magnetic diffusivity, $\sigma$ and $\mu_0$ being the electric conductivity of the fluid and the magnetic permeability of vacuum; and $\nu$ is the kinematic viscosity. At low $\Rm$ and $\Pm$, flows affected by an externally imposed magnetic field can be accurately described within the framework of the quasi-static approximation.\cite{davidson2017}  The perturbations of the magnetic field are much weaker than the imposed magnetic field and can be neglected in the expressions of Ohm's law and Lorentz force. The effect of the imposed magnetic field on the flow can be approximated as a one-way effect decoupled from the reverse effect of the flow on the magnetic field.

In this paper we consider a version of the Hartmann channel flow--an archetypal system for flows with MHD effect of a constant imposed magnetic field.\cite{Hartmann1937a,Hartmann1937b} An electrically conducting fluid is driven by an imposed longitudinal pressure gradient between two electrcially insulating walls in the presence of the wall-normal magnetic field.
Two nondimensional parameters determine the flow state: the hydrodynamic Reynolds number $\Rey\equiv Uh/\nu$ and the Hartmann number $\Ha=Bh\sqrt{\sigma / \rho \nu}$, where $B$ is the induction of the applied magnetic field and $\rho$ is the density.

Significant flow transformation is observed high $\Ha$. The magnetic field flattens the laminar velocity profile in the core of the channel and concentrates the viscous stresses within the Hartmann boundary layers, whose typical thickness scales as $\delta_{Ha} \sim h\Ha^{-1}$. 

The instability and transition to turbulence in the classical Hartmann channel flow are relatively well understood (see, e.g., Ref. \onlinecite{zikanov2014} for a review). The process is initiated within the Hartmann layers and occurs at higher $\Rey$ than in the hydrodynamic channel flow. In fact, it has been shown that at $\Ha\gg 1$, the instability is determined by the modified Reynolds number, which uses $\delta_{Ha}$ as the typical length scale: $\R\equiv U\delta_{Ha}/\nu=\Rey/\Ha$.    The most accurate estimate of the linear stability limit is $\R_{cr,lin}=48250$.\citep{lingwood:1999}

As in other wall-bounded parallel shear flows, the actual transition to turbulence in the Hartmann flow occurs at Reynolds numbers much lower than the linear instability threshold. For example, a compilation of earlier experiments\cite{branover1978} gives the critical Reynolds number $\R_{cr}\approx 215$. Subsequent experiments and numerical simulations have shown that for $\Ha \gtrsim 20$, the transition occurs in the range $200 < \R < 400$.\cite{zikanov2014} The discrepancy with predictions from linear stability analysis is attributed, similarly to other wall-bounded parallel shear flows, to the nonlinear bypass transition mechanism. \citep{krasnov:2004}

It is, in general, evident that the variation of physical properties within the flow domain affects the stability properties of the flow. In the case of MHD flows in ducts and channels, the effect of electric conductivity is expected to be particularly strong because conductivity determines the distribution of electric currents, which determines the distribution of the Lorentz force that, in turn, strongly influences the base velocity profile. To date, this effect has been demonstrated for variations in the electrical conductivity of the walls. A well-known example is the so-called Hunt flow—the flow in a duct where the two walls parallel to the imposed magnetic field are electrically insulating, while the walls perpendicular to the magnetic field are electrically conducting. \citep{hunt:1965} In such a flow, the base velocity profile exhibits symmetric, quasi-planar high-speed jets near the insulating walls. A distinctive feature of these jets is the presence of inflection points in the velocity profile. Analyzes conducted for both infinitely large \citep{priede:2010} and finite \citep{arlt:2017} electrical conductivity of the field-normal walls show that the formation of these jets renders the flow linearly unstable, even at moderate values of the Hartmann number.

Another example of this effect is the Hartmann flow in a layer bounded by insulating plates, each containing a conducting strip oriented along the flow direction.\cite{buhler1996} Near the conducting sections of the boundaries, the flow velocity decreases forming inflection points in the spanwise velocity profile. This, in turn, gives rise to instabilities characterized by the formation of time-dependent, vortex-type flow patterns.\cite{buhler1996}

In this paper, we consider a different type of variation of physical properties, namely the variation of the electric conductivity and viscosity of the fluid itself. The recently published generalization of the Hartmann solution\cite{Okatev2023} shows that the variations, most notably that of the electric conductivity, lead to a strong deformation of the base laminar profile of velocity. It has been shown that the profile may develop inflection points. Here, we analyze the consequences of this transformation for the stability of the flow. As the first step of the analysis, we focus on linear instability. The role of the nonlinear bypass mechanism and the effects related to the density change and the appearance of buoyancy forces are left to future studies.

\section{Problem setting and base flow state}
\label{sec:basic_flow}

\begin{figure*}
    \centering
    \includegraphics[width=10cm]{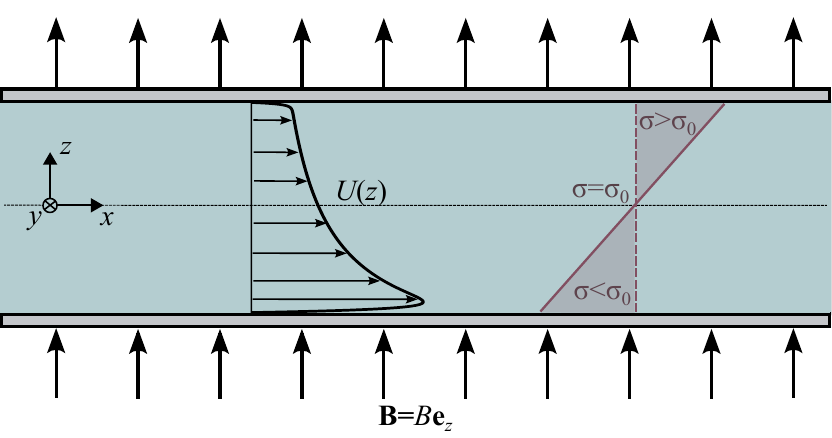}
    \caption{Schematics of the flow. {Variation of the electric conductivity of the fluid $\sigma$ and the corresponding profile of the base flow velocity $U(z)$ are illustrated.}}
    \label{fig:scheme}
\end{figure*}

We consider the flow of an electrically conducting,  incompressible, viscous Newtonian fluid in a channel with non-conducting walls. A wall-normal magnetic field of induction $B$ is imposed (see figure \ref{fig:scheme}). Streamwise, spanwise, and wall-normal coordinates are denoted by $x$, $y$, and $z$, respectively. The kinematic viscosity $\nu$ and the electrical conductivity $\sigma$ of the fluid may vary.  The MHD equations in the quasistatic approximation can be written in terms of the velocity $\mathbf{u}$, pressure $p$, electric current density $\mathbf{j}$ and electric potential $\varphi$. Using the mean streamwise velocity $U$ and a half-width of the channel $h$ as typical velocity and length scales, and $h/U$, $B$, $\sigma_0 U B$ and $h U B$ as typical scales of time, magnetic field, electric current, and electric potential, respectively, the non-dimensional governing equations and boundary conditions are written in form
\begin{eqnarray}
    \frac{\partial \mathbf{u}}{\partial t} + \mathbf{u}\cdot\nabla\mathbf{u} & = & -\nabla p + \frac{1}{\Rey}\nabla\cdot2\nu\mathbf{S} + \frac{\Ha^2}{\Rey}\mathbf{j}\times\mathbf{e}_z,\label{eq:NS}\\
    \nabla \cdot \mathbf{u} & = & 0,\\
    \mathbf{j} & = & \sigma\left(-\nabla \varphi + \mathbf{u}\times\mathbf{e}_z\right),\label{eq:ohm}\\
    \nabla \cdot \left(\sigma \nabla\varphi\right) & = & \nabla\cdot\left(\sigma\mathbf{u}\times\mathbf{e}_z\right), \label{eq:potential}\\
    \mathbf{u}=\mathbf{0}, ~ \frac{\partial \varphi}{\partial n} & = & 0 \quad \mathrm{at} \quad z=\pm1, \label{eq:bc}
\end{eqnarray}
where $\mathbf{S}$ is the rate of strain tensor. The Hartmann number $\Ha = B h \sqrt{\sigma_0 / \rho \nu_0}$ and the Reynolds number $\Rey = U_0 h / \nu_0$ are defined using the mean values of viscosity $\nu_0$ and conductivity $\sigma_0$.

In this paper, we do not consider the transport of the variable coefficients $\nu$ and $\sigma$ by the flow. Fixed stationary distributions $\nu(z)$ and $\sigma(z)$ corresponding to the base state are assumed. This simplification is consistent with the main focus of this paper on the mechanisms of linear instability and, therefore, on the dynamics of small-amplitude perturbations, which do not significantly modify $\nu(z)$ and $\sigma(z)$. Linear distributions with constant slopes $\varkappa$, $\varsigma$, and the mean values corresponding to the values at $z=0$ are used:
\begin{equation}
    \sigma(z) = 1 + \varkappa z, \quad \nu(z) = 1 + \varsigma z. \label{eq:linear_var}
\end{equation}
Our motivation for using the linear profiles is as follows. The linear profile is, by far, the simplest possible dependence. Furthermore, it is an acceptable approximation for layers with an imposed \emph{stable} temperature or admixture gradient. In particular, the electrical conductivity of metals decreases rapidly with increasing temperature. Thus, the flow of a liquid metal layer heated from above and subjected to a vertical magnetic field can be considered a prototype of the system studied here.

Using \eqref{eq:NS} and \eqref{eq:ohm} we can obtain the equation for the stationary base velocity profile $U(z)$
\begin{equation}
    \frac{d^2U}{dz^2} = -\frac{\Rey}{\nu}\frac{d p}{d x} + \Ha^2\left(\frac{\sigma}{\nu}\right)U - \frac{1}{\nu}\frac{d \nu}{dz}\frac{d U}{d z}.
    \label{eq:statp1}
\end{equation}
The pressure gradient is determined by the normalization condition
\begin{equation}
    \frac{1}{2}\int_{-1}^{1}U(z)dz = 1.
    \label{eq:normc}
\end{equation}
With linear distributions of $\sigma$ and $\nu$,  \eqref{eq:statp1} becomes
\begin{equation}
    \frac{d^2U}{dz^2} = -\frac{\Rey}{1 + \varsigma z}\frac{d p}{d x} + \Ha^2\left(\frac{1 + \varkappa z}{1 + \varsigma z}\right)U - \frac{\varsigma}{1 + \varsigma z}\frac{d U}{d z}.
    \label{eq:statp2}
\end{equation}
We note that, due to the symmetry of the equations $\left\{\varkappa\rightarrow-\varkappa, \varsigma\rightarrow-\varsigma, z\rightarrow-z\right\}$, it is sufficient to consider only positive values of $\varkappa$ and $\varsigma$. Solutions for negative values can be obtained by reflecting the resulting velocity profiles about the $x-y$ plane.

\begin{figure}
    \centering
    \includegraphics[height=4.4cm]{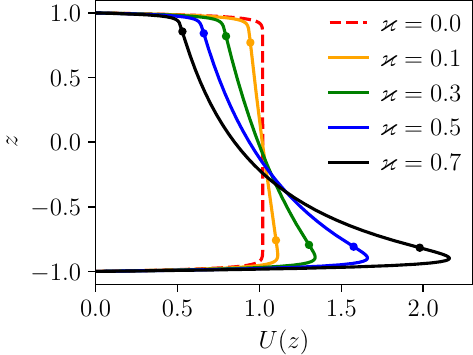}(a)
    \includegraphics[height=4.4cm]{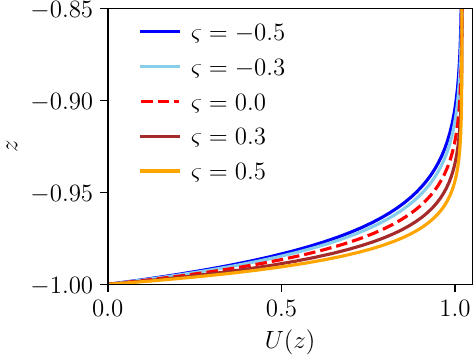}(b)
    \caption{\textit{(a)} Base velocity profiles at $\Ha = 50$, $\varsigma = 0$ under varying conductivity.  {Inflection points are indicated by dots. \textit{(b)} Base velocity profiles within the boundary layer at $\Ha=50$, $\varkappa=0$ under varying viscosity.} Red dashed line corresponds to the ordinary Hartmann flow profile.}
    \label{fig:ubase}
\end{figure}

\begin{figure}
    \centering
    \includegraphics[height=4.5cm]{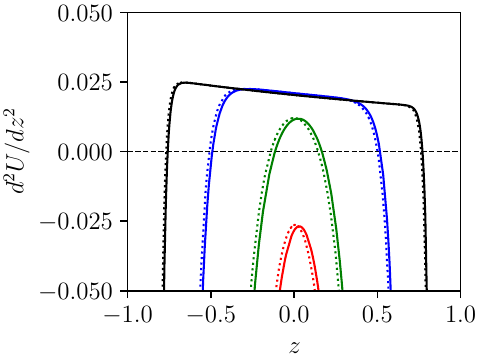}(a)
    \includegraphics[height=4.5cm]{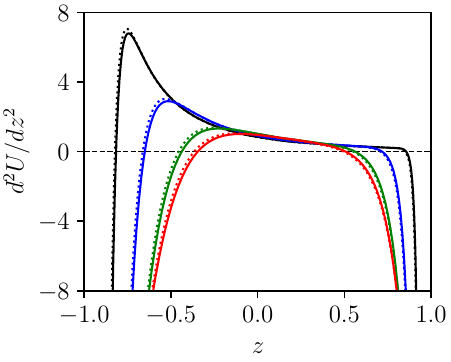}(b)
    \caption{$d^2U/dz^2$ at $\Ha=8$ (red), $\Ha=10$ (green), $\Ha=20$ (blue), $\Ha=50$ (black) for $\varkappa=0.1$ \textit{(a)} and $\varkappa=0.7$ \textit{(b)}. Solid lines correspond to uniform viscosity ($\varsigma=0$) and dotted lines correspond to $\varsigma=0.1$.}
    \label{fig:ip_1}
\end{figure}

The numerical solution  {of \eqref{eq:statp2}} is presented and discussed in Ref. \onlinecite{Okatev2023}. The most interesting results are obtained for the case of variable electric conductivity ($\varkappa>0$). The velocity profile becomes asymmetric and, at sufficiently high $\Ha$ and $\varkappa$, develops inflection points (see figure \ref{fig:ubase}a).  {Development of inflection points is further illustrated in figure \ref{fig:ip_1}, where we show numerically calculated $d^2U/dz^2$ at various $\Ha$ and $\varsigma$  {for $\varkappa=0.1$ and $\varkappa=0.7$.}}  

 While obtaining exact analytical solutions for \eqref{eq:statp1} is complicated, even for linear distributions of $\sigma$ and $\nu$, it is possible to make some arguments about the appearance of inflection points based on simple asymptotic behavior considerations. At $\Ha\gg1$ and $\varkappa\ll1$ the velocity gradients in the core of the flow are small, so the effect of viscosity can be neglected. The velocity profile outside the boundary layers can therefore be approximated as defined by a balance between the Lorenz force and the imposed pressure gradient:
\begin{equation}
    \Rey\frac{dp}{dx} = \Ha^2\sigma(z) U(z)    
\end{equation}
from which it follows that $U(z)\sim1/{\sigma(z)}$. Assuming $\varsigma \ll 1 $, and incorporating boundary layers by analogy with Ref. \onlinecite{Lock1955}, we find that for our case with a linear distribution of properties,
\begin{equation}
\begin{split}
    U(z) \sim \frac{1}{1+\varkappa z} - \frac{\exp\left[-\Ha\sqrt{\frac{1-\varkappa}{1-\varsigma}}(1+z)\right]}{1-\varkappa} \\ - \frac{\exp\left[-\Ha\sqrt{\frac{1+\varkappa}{1+\varsigma}}(1-z)\right]}{1+\varkappa}.
\end{split}
\end{equation}
The inflection points are defined as the zeros of 
\begin{equation}
\begin{split}
    U''(z) \sim \frac{2\varkappa^2}{(1+\varkappa z)^3}-\frac{\Ha^2\exp\left[-\Ha\sqrt{\frac{1-\varkappa}{1-\varsigma}}(1+z)\right]}{1-\varsigma} \\ - \frac{\Ha^2\exp\left[-\Ha\sqrt{\frac{1+\varkappa}{1+\varsigma}}(1-z)\right]}{1+\varsigma}.
    \label{eq:d2u_approx}
\end{split}
\end{equation}
At $\varkappa=0$, $U''(z)$ does not change sign; therefore, the linear variation of viscosity by itself cannot create inflection points.

\begin{figure}
    \centering
    \includegraphics[height=5cm]{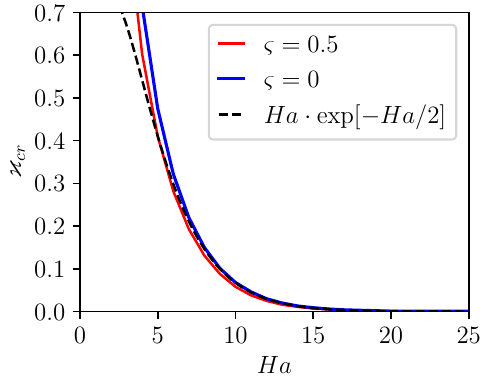}
    \caption{Numerically calculated $\varkappa_{cr}$, such that inflection points do not appear at $\varkappa<\varkappa_{cr}$, but appear at $\varkappa>\varkappa_{cr}$, versus $\Ha$ for two values of $\varsigma$. Black dashed line corresponds to the asymptotic relation \eqref{eq:kap_crit}.}
    \label{fig:kappa_crit}
\end{figure}

As Ha increases, inflection points appear at increasingly smaller values of $\varkappa$ in the vicinity of $z=0$.  Considering $\varkappa\rightarrow0$ and $\varsigma=0$, we obtain from \eqref{eq:d2u_approx} an  asymptotic expression
\begin{eqnarray}
    U''(0) \sim 2\varkappa^2 - 2\Ha^2\exp\left[-\Ha\right]
\end{eqnarray}
and find the critical value $\varkappa_{cr}(\Ha)$, such that inflection points are absent at $\varkappa < \varkappa_{cr}$ but present at $\varkappa > \varkappa_{cr}$:
\begin{equation}
    \varkappa_{cr}\approx\Ha\exp\left[-Ha/2\right].
    \label{eq:kap_crit}
\end{equation}
As one can see in figure \ref{fig:kappa_crit} this approximation agrees fairly well with the numerical results for $\Ha\gtrsim 10$.  $\varkappa_{cr}(\Ha)$ decreases rapidly with $\Ha$ and becomes smaller than $10^{-3}$ at $\Ha>20$.

The effect of the variation in viscosity ($\varsigma >0$) is less dramatic. As shown in figure \ref{fig:ubase}b, it changes the thickness of the boundary layers, which may affect the stability of the flow. 
It also influences the threshold for the appearance of inflection points (see figure \ref{fig:ip_1}), but this effect is not strong and weakens with increasing $\Ha$ due to the growing dominance of the Lorenz force (see figure \ref{fig:kappa_crit}). Since relatively large values of $\Ha$ that are interesting for applications are considered in the rest of this paper, we assume $\varsigma=0$ in the further analysis.

\section{Linear stability analysis}\label{sec:linear}

The presence of inflection points suggests the possibility of linear instability associated with them. This would represent a new instability mechanism in the Hartmann channel flow. The linear stability analysis exploring this possibility is carried out in this section.

We assume uniform viscosity $\varsigma=0$ and focus the analysis on the effect of linearly varying conductivity \eqref{eq:linear_var} with $\varkappa$ between 0 (the classical Hartmann channel flow) and 0.8. The analysis is performed at $\Ha=25$, 50, 100.

Linear stability of the base flow $U(z)$ to infinitesimal perturbations $\mathbf{u}$, $p$ and $\varphi$ of velocity, pressure, and electric potential is analyzed. The linearized governing equations are transformed, following the standard procedure,\cite{Schmid2001} into scalar equations for the $z$-component of velocity $u_z$ and vorticity $\xi=\partial u_y/\partial x - \partial u_x/\partial y$, and potential $\varphi$:
\begin{small}
\begin{gather}
    \left[\left(\frac{\partial}{\partial t} + U \frac{\partial}{\partial x}\right)\Delta - \frac{d^2 U}{d z^2} \frac{\partial}{\partial x}\right]u_z = \frac{1}{\Rey}\Delta \Delta u_z - \frac{\Ha^2}{\Rey}\left(\varkappa\frac{\partial u_z}{\partial z} + \sigma\frac{\partial^2 u_z}{\partial z^2} \right),
    \label{eqw_orr}\\
    \left(\frac{\partial}{\partial t} + U \frac{\partial}{\partial x}\right)\xi = \frac{1}{\Rey}\Delta\xi + \frac{d U}{d z}\frac{\partial u_z}{\partial y} - \frac{\Ha^2}{\Rey}\left(\varkappa\frac{\partial \varphi}{\partial z} + \sigma\frac{\partial^2 \varphi}{\partial z^2} \right),
    \label{eqeta_sq}\\
    \Delta \varphi + \frac{\varkappa}{\sigma}\frac{\partial \varphi}{\partial z} = \xi.
\end{gather}
\end{small}
Considering wavelike perturbations of the form
\begin{equation}
\small{
     (u_z,\xi,\varphi)(x,y,z,t) = (\hat{w},\hat{\xi},\hat{\varphi})(z)\exp\left\{\mathrm{i}\left(\alpha x + \beta y - \gamma t \right) \right\},
}
\end{equation}
where $\alpha$ and $\beta$ are the real-valued wavenumbers, and $\gamma=\gamma_r+ \mathrm{i} \gamma_i$ is the eigenvalue, we obtain the Orr-Sommerfeld and Squire equations modified for the presence of the Lorentz force:
\begin{small}
\begin{eqnarray}\label{eq:OS}
\begin{split}
    \left(U - c\right)\left(\hat{w}'' - k^2 \hat{w}\right) - U'' \hat{w} & = \frac{1}{\mathrm{i}\alpha\Rey}\left(\hat{w}^{IV} - 2 k^2\hat{w}'' + k^4 \hat{w}\right) \\ & - \frac{\Ha^2}{\mathrm{i}\alpha\Rey}\left(\varkappa\hat{w}' + \sigma\hat{w}'' \right)
\end{split}\\
    \left(U - c\right)\hat{\xi} = \frac{1}{\mathrm{i} \alpha \Rey}\left(\hat{\xi}'' - k^2 \hat{\xi}\right) + \frac{\beta}{\alpha}U'\hat{w} - \frac{\Ha^2}{\mathrm{i} \alpha \Rey}\left(\varkappa\hat{\varphi}' + \sigma\hat{\varphi}'' \right).
\end{eqnarray}
\end{small}
Here $c = \gamma / \alpha$ and $k^2 = \alpha^2 + \beta^2$. The problem statement is completed by the electric potential equation
\begin{equation}
\small{
    \hat{\varphi}'' - k^2 \hat{\varphi} + \frac{\varkappa}{\sigma}\hat{\varphi}' = \hat{\xi}
}
\label{eq:phispec}
\end{equation}
and the boundary conditions
\begin{small}
\begin{gather}
    \hat{w}(-1) = \hat{w}(1) = 0, \ \  \hat{w}'(-1) = \hat{w}'(1) = 0,\\   
    \hat{\xi}(-1) = \hat{\xi}(1) = 0, \ \
    \hat{\varphi}'(-1) = \hat{\varphi}'(1) = 0.
\end{gather}
\end{small}

Since the Squire's theorem is valid in the case of a channel flow with wall-normal magnetic field,\citep{Takashima1996, lingwood:1999} $\beta=0$ is assumed in the linear stability analysis in the remainder of this section.

The instability parameters are determined from the numerical solution of the Orr-Sommerfeld equation \eqref{eq:OS} performed using the spectral code Dedalus.\cite{Burns2020} Chebyshev polynomial expansion is used. The convergence of the solution with the number of modes used in the spectral decomposition is shown in Table \ref{tab:convergence} for both the Orr-Sommerfeld equation \eqref{eq:OS} and the full spectral problem \eqref{eq:OS}-\eqref{eq:phispec}. For all the calculations reported in rest of this section, $N=384$ is used. The key outcome of the solution is the eigenvalue with the largest $\gamma_i$, which determines stability.

\begin{table}
\caption{\label{tab:convergence} Eigenvalues $c$ with the largest imaginary part calculated using decomposition over $N$ Chebyshev polynomials. All calculations performed with $\Ha=50$, $\Rey=25000$, $\alpha=\pi/3$, $\beta=0$, $\varkappa=0.7$.}
\begin{ruledtabular}
\begin{tabular}{c|c|c}
$N$ & $c$, Orr-Sommerfeld & $c$, Orr-Sommerfeld-Squire\\
\hline
48 & 1.9055151103+0.013998877$\mathrm{i}$ & 1.9055151103+0.013998877$\mathrm{i}$\\
64 & 1.9054773186+0.014278013$\mathrm{i}$ & 1.9054773186+0.014278013$\mathrm{i}$\\
96 & 1.9054663707+0.014275107$\mathrm{i}$ & 1.9054663707+0.014275107$\mathrm{i}$\\
128 & 1.9054664185+0.014275090$\mathrm{i}$ & 1.9054664185+0.014275090$\mathrm{i}$\\
192 & 1.9054664051+0.014275092$\mathrm{i}$ & 1.9054664051+0.014275092$\mathrm{i}$\\
256 & 1.9054663719+0.014275087$\mathrm{i}$ & 1.9054663720+0.014275087$\mathrm{i}$\\
384 & 1.9054663810+0.014275092$\mathrm{i}$ & 1.9054663810+0.014275092$\mathrm{i}$\\
512 & 1.9054663916+0.014275107$\mathrm{i}$ & 1.9054663918+0.014275107$\mathrm{i}$\\
768 & 1.9054663820+0.014275090$\mathrm{i}$ & 1.9054663822+0.014275090$\mathrm{i}$\\
\end{tabular}
\end{ruledtabular}
\end{table}

For each explored set $(\Ha,\varkappa,\Rey)$, solutions are obtained in a range of $\alpha$. The critical values $\Rey_{cr}(\Ha,\varkappa)$ and $\alpha_{cr}(\Ha,\varkappa)$ are found using an iterative bisection-like algorithm. It operates in a predetermined range of wavenumbers $\alpha \in \left[\alpha_1, \alpha_2\right]$. Iterations start with $\Rey_{min}$, such that $\gamma_i<0$ at $\alpha \in \left[\alpha_1, \alpha_2\right]$  and $\Rey_{max}$, such that $\gamma_i$ has two zeros $\alpha_1^*$ and $\alpha_2^*$ in $\alpha \in \left[\alpha_1, \alpha_2\right]$.
Bisection iterations continue until $\Rey_{cr}$ is found, at which $\gamma_i$ has two zeros satisfying $|\alpha_2^* - \alpha_1^*| < 10^{-6}$. The  {critical} wavenumber is evaluated as $\alpha_{cr}=(\alpha_1^* + \alpha_2^*)/2$.

The solution procedure was verified by repeating the known results for the hydrodynamic Poiseuille flow \citep{Schmid2001} and Hartmann channel flow.\cite{lingwood:1999} Furthermore, selected solutions were confirmed by directly solving the linearized version of \eqref{eq:NS}-\eqref{eq:bc} using the finite-difference method described in section \ref{sec:num}.

\begin{figure*}
    \centering
    \includegraphics[width=0.9\linewidth]{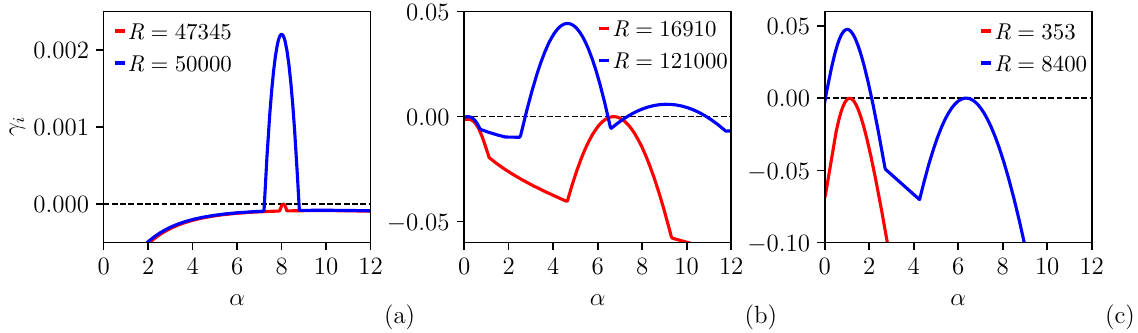}
    \caption{ {Dispersion curves at $\Ha=50$ for $\varkappa=0$ (a),  $\varkappa = 0.5$ (b), $\varkappa = 0.7$ (c). Peaks at low wavelengths ($\alpha<3$) correspond to the inflection point instability. Peaks at moderate and high wavelengths ($\alpha>3$) correspond to the Hartmann layer instability.}}
    \label{fig:growthrate}
\end{figure*}

\begin{figure*}
    \centering
    \includegraphics[height=4.5cm]{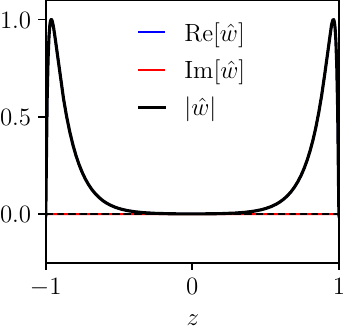}(a)
    \includegraphics[height=4.5cm]{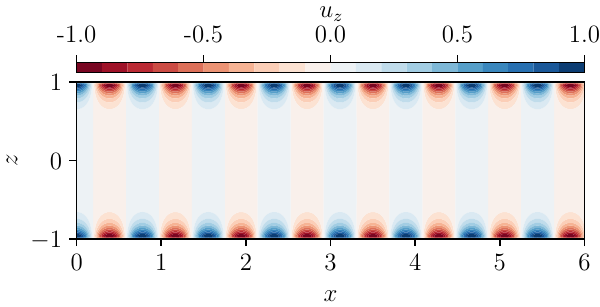}(b)
    \caption{ \textit{(a)} Fastest growing eigenmodes of Orr-Sommerfeld equation perturbations at $\Ha=50$, $\varkappa=0$, $\Rey = 2.5\times10^6$, $\alpha = 8$. \textit{(b)}  Corresponding vertical velocity fields.}
    \label{fig:eigenmodes_k00}
\end{figure*}

\begin{figure*}
    \centering
    \includegraphics[height=4.5cm]{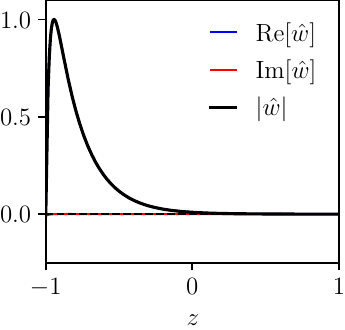}(a)
    \includegraphics[height=4.5cm]{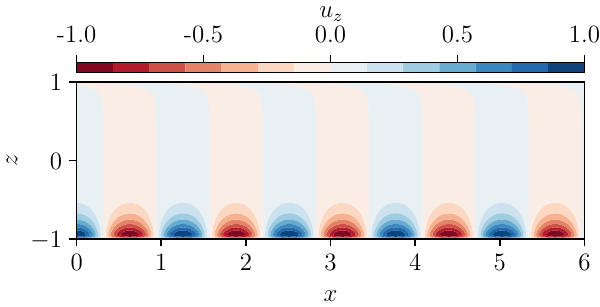}(b)\\
    \includegraphics[height=4.5cm]{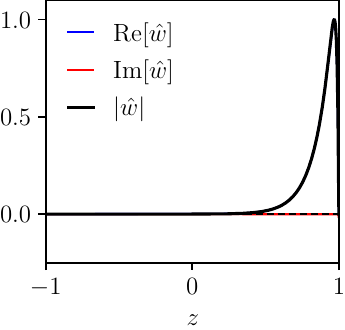}(c)
    \includegraphics[height=4.5cm]{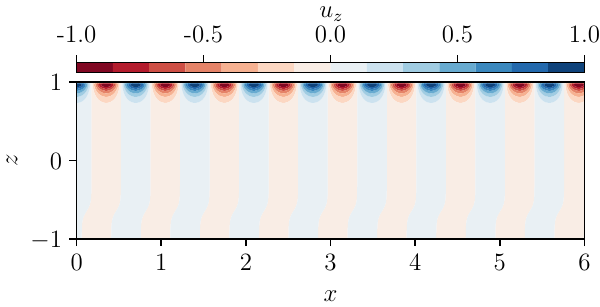}(d)
    \caption{\textit{(a)} Fastest growing eigenmodes of Orr-Sommerfeld equation at $\Ha=50$, $\varkappa=0.5$, $\Rey = 6.05\times10^6$. \textit{(b)}  Corresponding vertical velocity fields. Here $\alpha = 5$ \textit{(a,b)} and $\alpha = 9$ \textit{(c,d)}.}
    \label{fig:eigenmodes_k05}
\end{figure*}

Two types of instability are found. One is \emph{Hartmann Layer Instability (HLI)} illustrated in figures \ref{fig:growthrate}--\ref{fig:eigenmodes_k05}. It appears at $\varkappa=0$ and $\varkappa>0$. As illustrated in figures \ref{fig:eigenmodes_k00} and \ref{fig:eigenmodes_k05}, the fastest growing eigenmodes are localized in the Hartmann boundary layers and are characterized by large values of $\alpha$.

At $\varkappa=0$, the instability threshold is $\Rey_{cr}=47345\Ha$ at $\alpha_{cr}=0.1615\Ha\approx 1/(6\delta_{Ha})$ in good agreement with the results reported earlier.\cite{Lock1955, Takashima1996, lingwood:1999} At $\Rey>\Rey_{cr}$, unstable modes are found in a narrow range of $\alpha$ (see figure \ref{fig:growthrate}a).  

The characteristics of HLI instability at $\varkappa>0$  are different from those at $\varkappa = 0$. Due to the asymmetry of the base velocity profile, the associated eigenmodes are asymmetric. As illustrated in figure \ref{fig:ubase}, the local base flow velocity is higher near the wall at $z = -1$. Consequently, the critical Reynolds number corresponds to neutral eigenmodes with velocity perturbations concentrated within just one Hartmann boundary layer adjacent to this wall. The Hartmann boundary layer at $z = 1$ becomes unstable at higher values of $\Rey$. Interestingly, because the boundary layer at $z = 1$ is thinner, its instability is associated with perturbations of a smaller typical length scale, i.e., eigenmodes of higher values of $\alpha$. This is demonstrated in figure \ref{fig:growthrate}b, which shows two peaks in the dispersion curve at $\Rey/\Ha = 121{,}000$: one at $\alpha \approx 5$ (representing the fastest-growing eigenmodes near $z = -1$), and another at $\alpha \approx 9$ (representing the fastest-growing eigenmodes near $z = 1$). Figure \ref{fig:eigenmodes_k05} provides further illustration, showing that, for the same flow parameters, the growing eigenmodes are localized near $z = -1$ or $z = 1$ depending on the value of $\alpha$.

\begin{figure*}
    \centering
    \includegraphics[height=5.0cm]{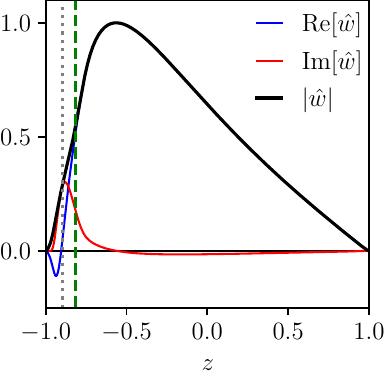}(a)
    \includegraphics[height=5.0cm]{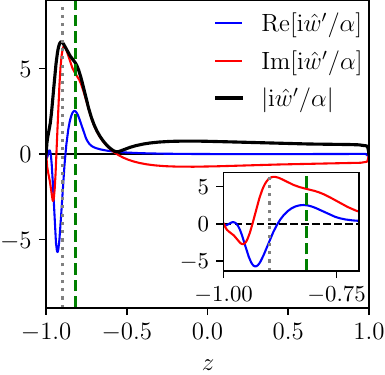}(b)\\
    \includegraphics[height=4.5cm]{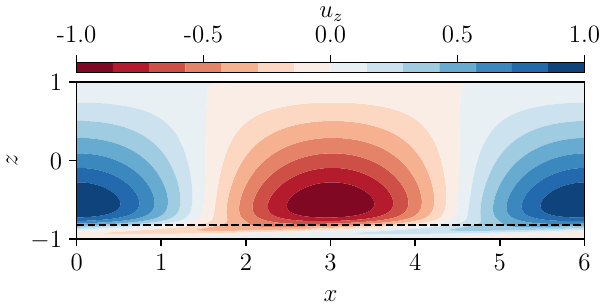}(c)\\
    \includegraphics[height=4.5cm]{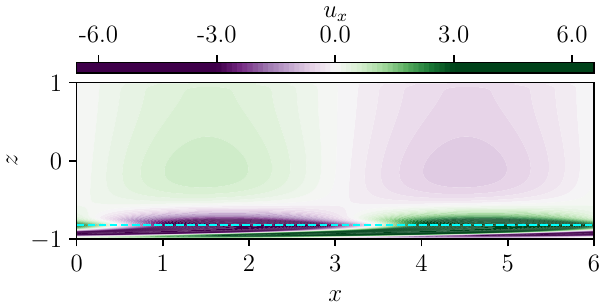}(d)\\
    \includegraphics[height=4.5cm]{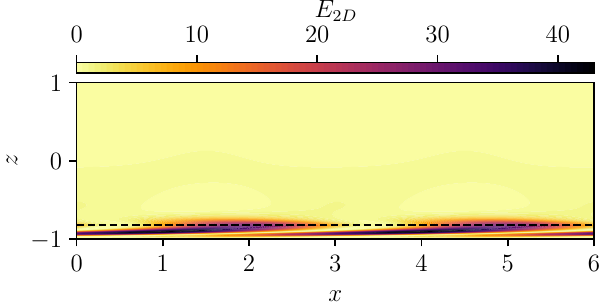}(e)
    \caption{ \textit{(a)} Fastest growing eigenmodes of the Orr–Sommerfeld equation at $\Ha=50$, $\varkappa=0.7$, $\Rey = 25{,}000$, and $\alpha = \pi/3$. \textit{(b)} Corresponding mode for the streamwise velocity perturbation, $\hat{u}x = \mathrm{i}\hat{w}'/\alpha$. \textit{(c, d, e)} Vertical and streamwise velocity fields, and distribution of perturbation energy, $E_{2D} = u_x^2 + u_z^2$. The dashed line corresponds to the location of the inflection point. The grey dotted line in \textit{(a, b)} indicates the location of the base velocity maximum.}
    \label{fig:eigenmodes_k07}
\end{figure*}

\begin{figure}
    \centering
    \includegraphics[height=5.cm]{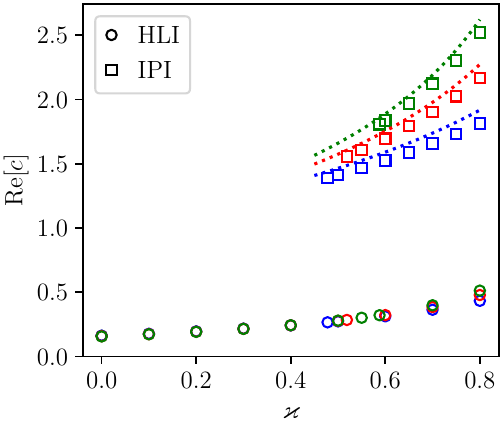}
    \caption{ {Real part of the phase velocity calculated for critical parameters $\{\Rey_{cr}, \alpha_{cr}\}$ at $\Ha=25$ (blue), $\Ha=50$ (red) and $\Ha=100$ (green) and shown as a function of $\varkappa$. Dotted lines correspond to the values of the base velocity at the inflection point.}}
    \label{fig:phase_vel}
\end{figure}

\begin{figure*}
    \centering
    \includegraphics[height=5.cm]{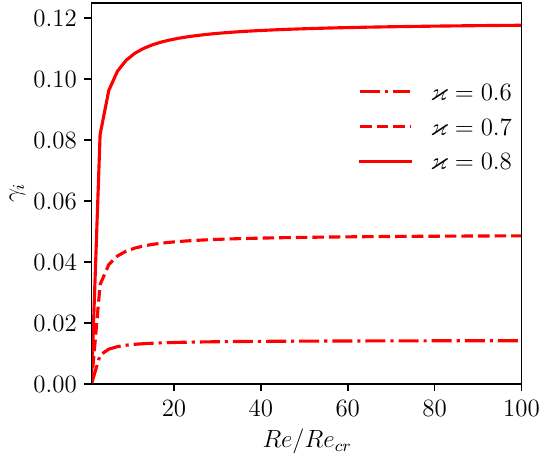} (a)
    \includegraphics[height=5.cm]{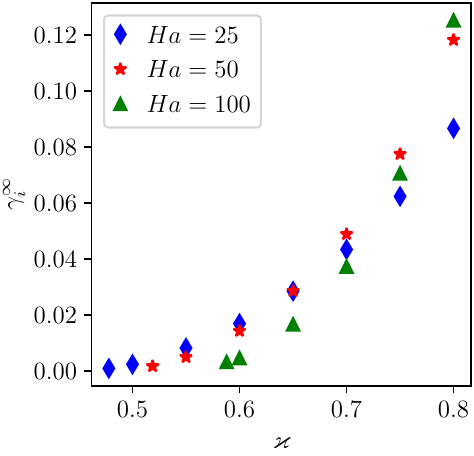} (b)
    \caption{  {\textit{(a)} Dependence of the IPI growth rate on $\Rey/\Rey_{cr}$ at $Ha=50$ for various $\varkappa$ calculated at the critical wavenumbers $\alpha_{cr}$. \textit{(b)} The limiting inviscid growth rate of IPI as a function of $\varkappa$ at $\Ha=25$ (blue), $\Ha=50$ (red) and $\Ha=100$ (green).}}
    \label{fig:gamma_infl}
\end{figure*}

\begin{figure*}
    \centering
    \includegraphics[height=4cm]{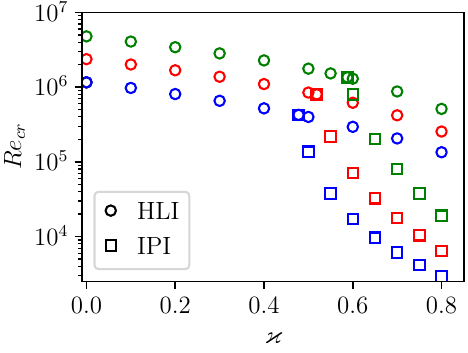}(a)
    \includegraphics[height=4cm]{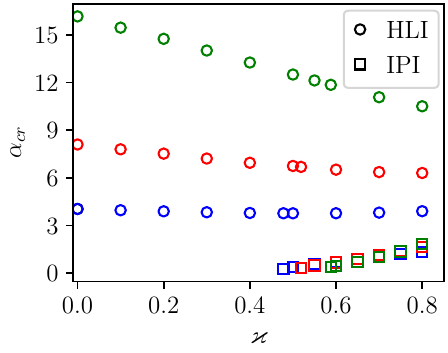}(b)\\
    \includegraphics[height=4cm]{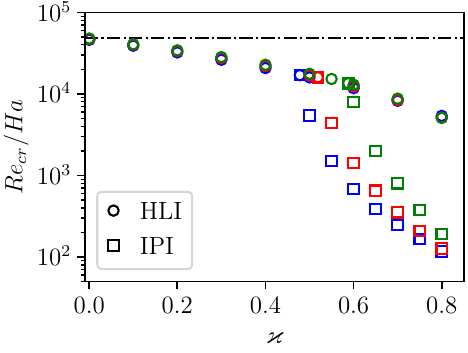}(c)
    \includegraphics[height=4cm]{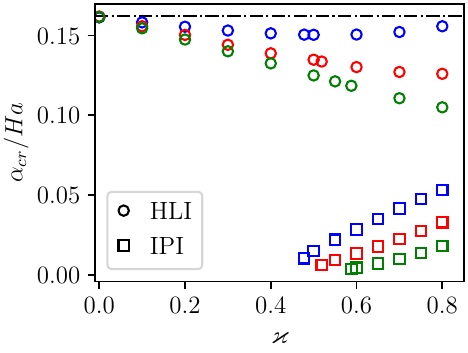}(d)
    \caption{Linear stability thresholds as functions of $\varkappa$ at $\Ha=25$ (blue), $\Ha=50$ (red) and $\Ha=100$ (green). \textit{(a)}, Critical Reynolds number $\Rey_{cr}$, \textit{(b)}, critical wavenumber $\alpha_{cr}$ (b), \textit{(c,d)}, Normalized values of $\Rey_{cr}$ and $\alpha_{cr}$.  {Black dash-dot line in \textit{(c, d)} corresponds to the linear instability of the classical Hartmann flow.\citep{lingwood:1999}}}
    \label{fig:linear_instability}
\end{figure*}

 Another type of instability appears only in flows at $\varkappa>0$ and, as we argue below, is likely to be associated with the development of inflection points in the base velocity profile. The unstable modes of this \textit{Inflection Point Instability (IPI)} are asymmetric and take the form of large spanwise-oriented rolls (see figures \ref{fig:eigenmodes_k07}). The instability is long-wave ($\alpha_{cr}<3$) across the entire explored range of $\varkappa$. 

The connection between the instability and the inflection points cannot be rigorously proven, as the classical inflection point instability is defined for inviscid flows without the MHD effect.\cite{Drazin2004} However, strong evidence suggesting such an association can be provided. One argument lies in the localization of the unstable eigenmodes near the inflection points. As shown in figure \ref{fig:eigenmodes_k07}, the streamwise velocity eigenmode—related to the Orr–Sommerfeld eigenmode $\hat{w}$ via $\hat{u}_x = \mathrm{i}\hat{w}/\alpha$—has its maximum $\mathrm{Re}[\hat{u}_x]$ very close to the inflection point (see figure \ref{fig:eigenmodes_k07}b). Furthermore, as demonstrated in figures \ref{fig:eigenmodes_k07}d,e, both the streamwise velocity perturbations $u_x$ and the total energy of the two-dimensional perturbations $E_{2D} = u_x^2 + u_z^2$ are localized in the vicinity of the inflection point. Interestingly, the maximum of $\mathrm{Im}[\hat{u}_x]$ coincides with the peak of the base velocity profile.

 A second argument is based on the behavior of the phase velocity, $c = \gamma/\alpha$ (see figure \ref{fig:phase_vel}). The calculations are presented for the critical values ${\Rey_{cr}, \alpha_{cr}}$. It can be seen that the real part of the phase velocity for the IPI correlates closely with the value of the base velocity profile at the inflection point. This observation is consistent with a scenario in which the instability develops around the inflection point, with the growing perturbations being advected by the mean flow.

Yet another argument supporting the similarity between the IPI mechanism and the inviscid mechanism of the classical inflection point instability is presented in figure \ref{fig:gamma_infl}a. As shown, the growth rate of the IPI has an inviscid limit. It approaches a limiting value as the Reynolds number increases, with this value nearly attained at $\Rey \approx 50 \Rey_{cr}$. This observation also allows us to calculate the limiting growth rates, $\gamma_i^{\infty}$, corresponding to the inviscid IPI mechanism for various values of $\Ha$ and $\varkappa$ (see figure \ref{fig:gamma_infl}b).

The described properties of the obtained solutions lead us to conclude that this type of instability is indeed related to the presence of an inflection point near the maximum velocity. However, we note that no unstable modes associated with the second inflection point near the upper wall (where the base flow velocity is lower) were found.

The separation in the wavenumber space allows us to calculate the thresholds $\Rey_{cr}$ of HLI and IPI separately. The results are summarized in figure \ref{fig:linear_instability}. We see that $\Rey_{cr}$ decreases with growing $\varkappa$ for both types of instability. HLI occurs at lower  $\Rey_{cr}$ than IPI at low and moderate values of $\varkappa$. In contrast, IPI occurs first at high $\varkappa$. The transition values $\varkappa_e$ such that  $\Rey_{cr}^{\text{HLI}}(\varkappa_e)=\Rey_{cr}^{\text{IPI}}(\varkappa_e)$ are listed in table \ref{tab:simult_inst} together with $\Rey_{cr}$ and the values of $\alpha_{cr}$ for both instabilities. 

\begin{table}
  \begin{center}
  \caption{Parameter values at which both instabilities appear simultaneously}
  \begin{ruledtabular}
  \begin{tabular}{lcccc}
       $\Ha$ & $\Rey_{cr}$& $\varkappa_e$ & $\alpha_{cr}^{\text{HLI}}$& $\alpha_{cr}^{\text{IPI}}$\\
       25   & ~~425000 & ~~0.4779~ & ~~3.76 & ~~0.26\\
       50   & ~~800800 & ~~0.5188 & ~~6.68 & ~~0.31\\
       100  & ~~1352000 & ~~0.5879 & ~~11.9 & ~~0.38\\
  \end{tabular}
  \end{ruledtabular}
  \label{tab:simult_inst}
  \end{center}
\end{table}

The typical situation at $\varkappa<\varkappa_e$ is illustrated in figure \ref{fig:growthrate}b for $\Ha=50$ and $\varkappa=0.5$. HLI appears at $\Rey=845500$, i.e., at $R=16910$. The maximum growth rate in the IPI wavenumber range remains negative until $\Rey=6050000$ ($R=121000$). We note that at such large values of $\Rey$ a third growing asymmetric mode appears corresponding to the instability of the Hartmann boundary layer near the upper wall  {(see figure \ref{fig:eigenmodes_k05}c,d)}.

An example of the situation at $\varkappa>\varkappa_e$ is shown in figure \ref{fig:growthrate}c for $\Ha=50$, $\varkappa=0.7$. We see that IPI appears at $\Rey=17665$ ($R=353.3$), while HLI first occurs at much larger $\Rey\approx420000$ ($R=8400$).

The data in figure \ref{fig:linear_instability}  provide information on possible scaling of $\Rey_{cr}$ and $\alpha_{cr}$. We see that the Hartmann layer scaling $\Rey_{cr}\sim \Ha$, $\alpha_{cr}\sim \Ha$ anticipated for HLI  {(see e.g. \cite{Lock1955, lingwood:1999})} at $\varkappa=0$ also applies at moderate $\varkappa>0$ (see figure \ref{fig:linear_instability}c,d. For IPI, the critical wavenumber $\alpha_{cr}$ is practically independent of $\Ha$ (see figure \ref{fig:linear_instability}b).

\section{Direct numerical simulations}
\label{sec:num}

The full system of governing equations \eqref{eq:NS}-\eqref{eq:bc} is solved numerically using the second-order finite-difference scheme shown to be accurate and effective for high-$\Ha$ MHD flows.\cite{Krasnov2011,Krasnov2023} The scheme is modified for the case of variable $\sigma$. Equations \eqref{eq:potential} and \eqref{eq:ohm} are written in terms of $f=\sigma \varphi$ and solved at each time step as
\begin{eqnarray}
    \label{eq:f1} \Delta f^{n+1} & = &\nabla\cdot\left(\sigma \mathbf{u}^{n+1}\times\mathbf{e}_z\right)+\nabla\cdot\left(\sigma^{-1}\tilde{f}^{n+1}\nabla\sigma\right),\\
    \label{eq:f2} \mathbf{j}^{n+1} & = & -\nabla f^{n+1}+\left(\sigma \mathbf{u}^{n+1}\times\mathbf{e}_z\right)+\sigma^{-1}\tilde{f}^{n+1}\nabla\sigma,
\end{eqnarray}
where $n$ is the time layer index and $\tilde{f}^{n+1}=2f^n-f^{n-1}$ is the second-order extrapolation from previous time layers.  The scheme retains the condition $\nabla \cdot \mathbf{j}^{n+1} = 0$. The extrapolation error is $\sim (\Delta t)^2$, i.e., of the same order as the error of the time-stepping algorithm of the scheme itself and is therefore tolerable. 

 The computational domain $[0, 6]\times[-L_y,L_y]\times[-1, 1]$ and the grid 128$\times N_y \times$128, clustered towards the walls, are used, with several values of $L_y$ and $N_y$ being tested in the solutions of linearized equations (see below) to detect possible three-dimensionality effects on the instability.  The computational grid is uniform along the $x$ (streamwise) and $y$ (spanwise) directions and clustered towards the $z$-walls according to the coordinate transformations $z = \tanh{(A\xi)} / \tanh{(A)}$. Here $A$ is the coefficient determining the degree of clustering, and $\xi$ is the transformed vertical coordinate, in which the grid is uniform. In all calculations, we set $A=2$ to accurately resolve the Hartmann layers.

Two kinds of simulations are completed. In one, linearized equations for perturbations of the base flow state are solved to verify the conclusions of the linear stability analysis. Simulations are performed starting with random three-dimensional noise and continued until the growth rate of the most unstable mode can be accurately determined. Solutions obtained at several sets of $(\Rey, \Ha, \varkappa)$ confirm the two-dimensional nature of the fastest growing perturbations and produce growth rates within 1\% of those predicted by the linear stability analysis of section \ref{sec:linear}.

Fully nonlinear simulations of the inflection point instability and the resulting transition to turbulence are presented here for $\Ha=50$, $\Rey=25000=1.415\Rey_{cr}^{\mathrm{IPI}}$, and $\varkappa=0.7$. The computational domain $[0, 6]\times[-1, 1]\times[-1, 1]$, grid 128$\times$64$\times$128, and time step $\Delta t=10^{-3}$ are used. The simulation starts with three-dimensional random velocity perturbations of amplitude $10^{-2}$ superimposed on the base flow and continues until a fully developed turbulent state is established.

It should be emphasized that these simulations serve as a demonstration of the bifurcation predicted by the linear stability analysis but do not capture the actual nonlinear evolution of the flow. This is because the linear distribution of $\sigma$ is assumed to remain unchanged throughout the transition, which does not reflect the true flow physics once the amplitude of the perturbations is no longer small and the cross-channel transport of the scalar field (such as temperature or admixture), responsible for the variation in $\sigma$, becomes significant. As we discuss in the concluding section of this paper, simulations of true nonlinear behavior will be an interesting topic for future research.

\begin{figure*}
    \centering
    \includegraphics[height=5.5cm]{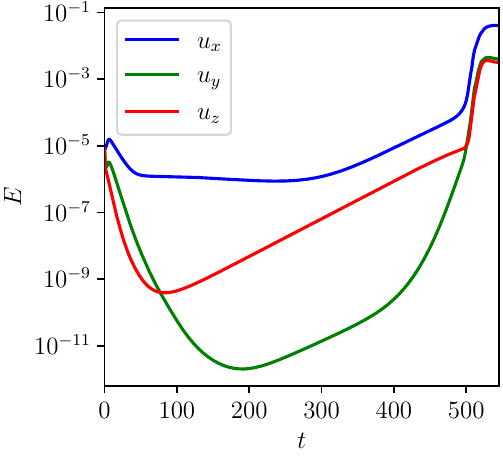}(a)
    \includegraphics[height=5.5cm]{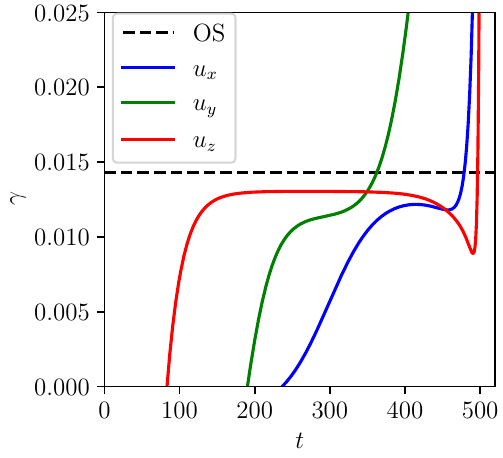}(b)    
    \caption{DNS of transition to turbulence at $\Ha=50$, $\Rey=25000$, $\varkappa=0.7$. \textit{(a)}, Volume-averaged kinetic energy associated with components of velocity perturbations, \textit{(b)}, Growth rate computed for each component. Black dashed line in \textit{(b)} corresponds to the growth rate predicted by linear stability analysis.}
    \label{fig:dns_enrg}
\end{figure*}

\begin{figure*}
    \centering
    \includegraphics[width=0.45\linewidth]{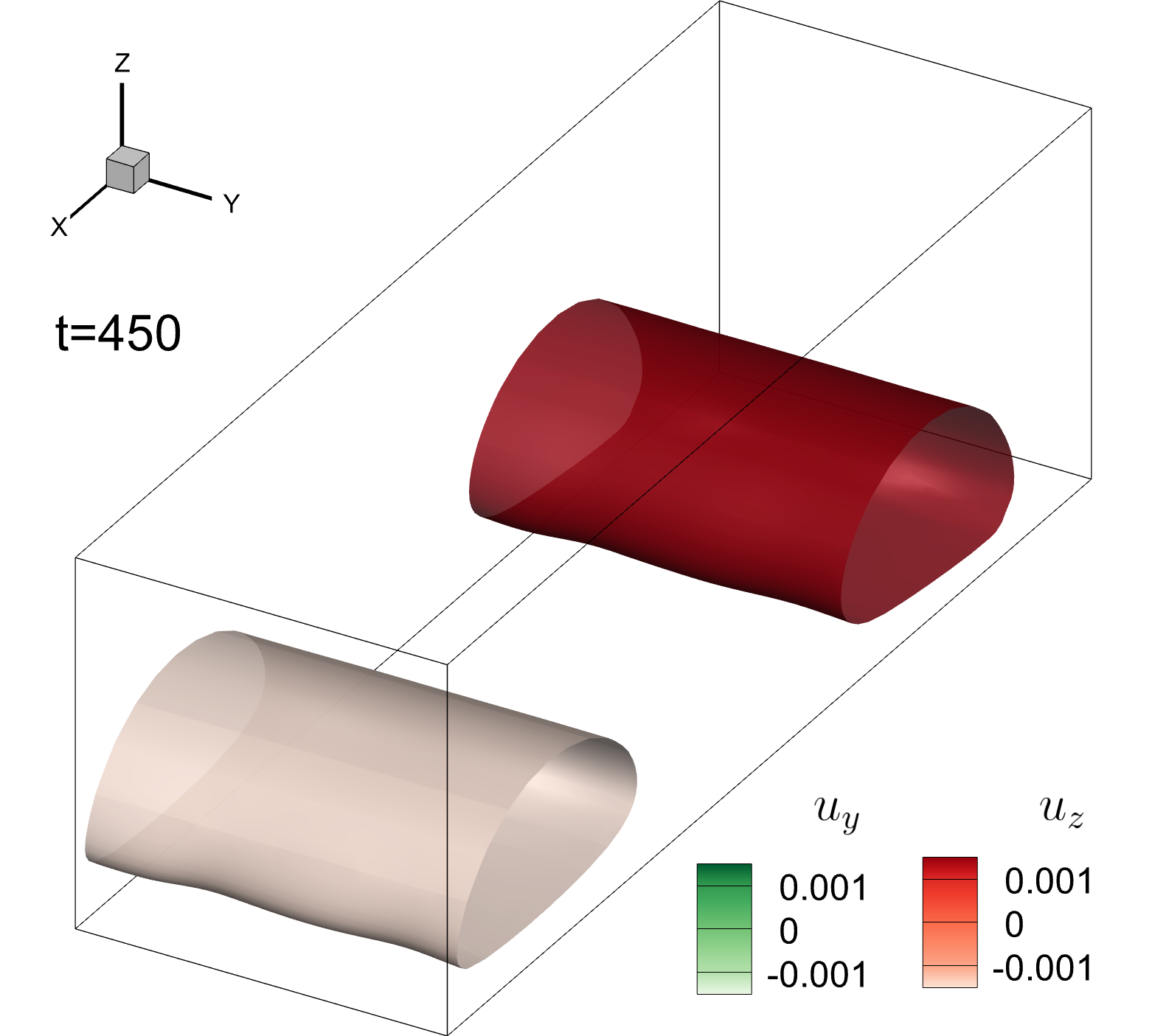}
    \includegraphics[width=0.45\linewidth]{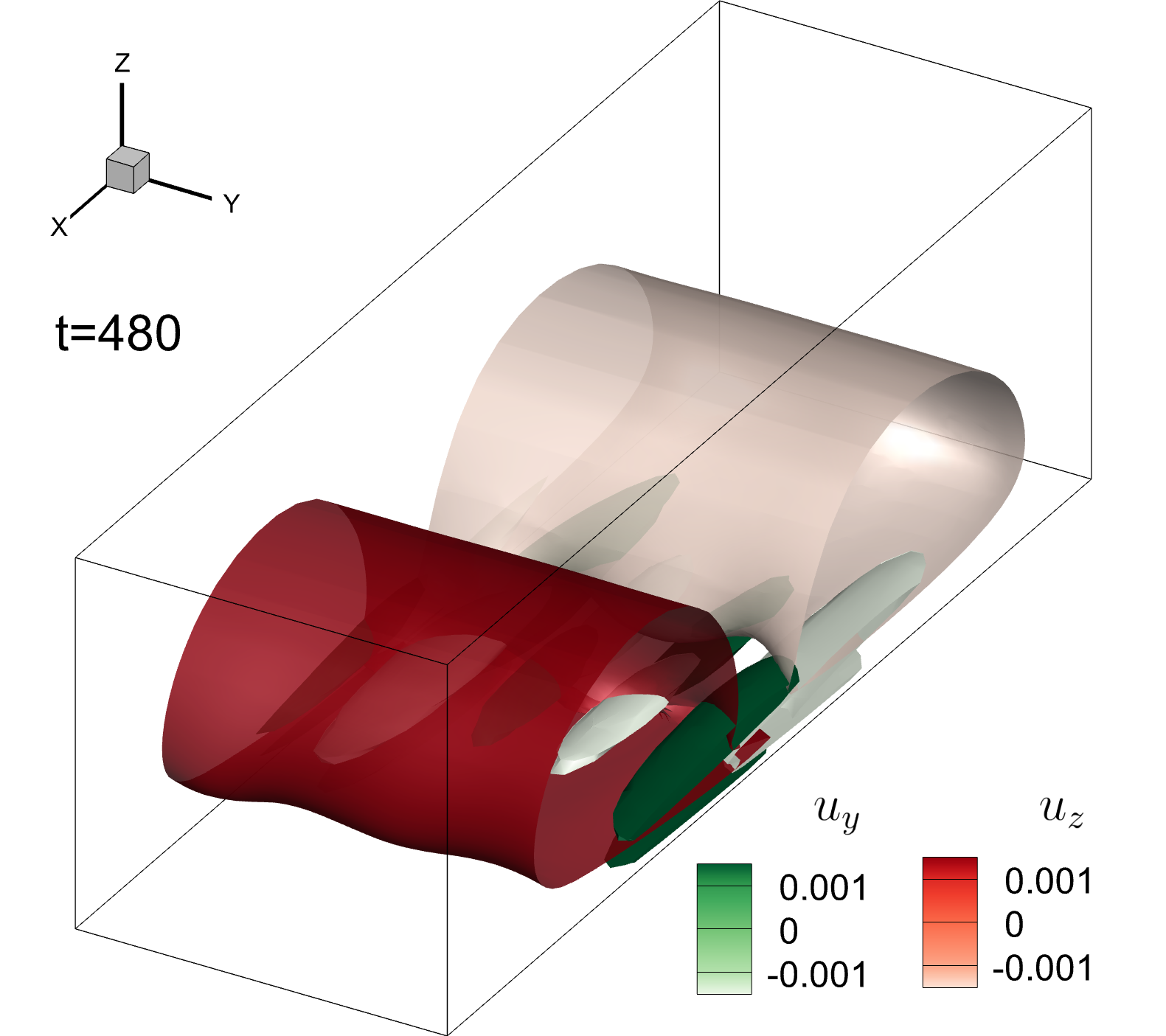}\\
    \includegraphics[width=0.45\linewidth]{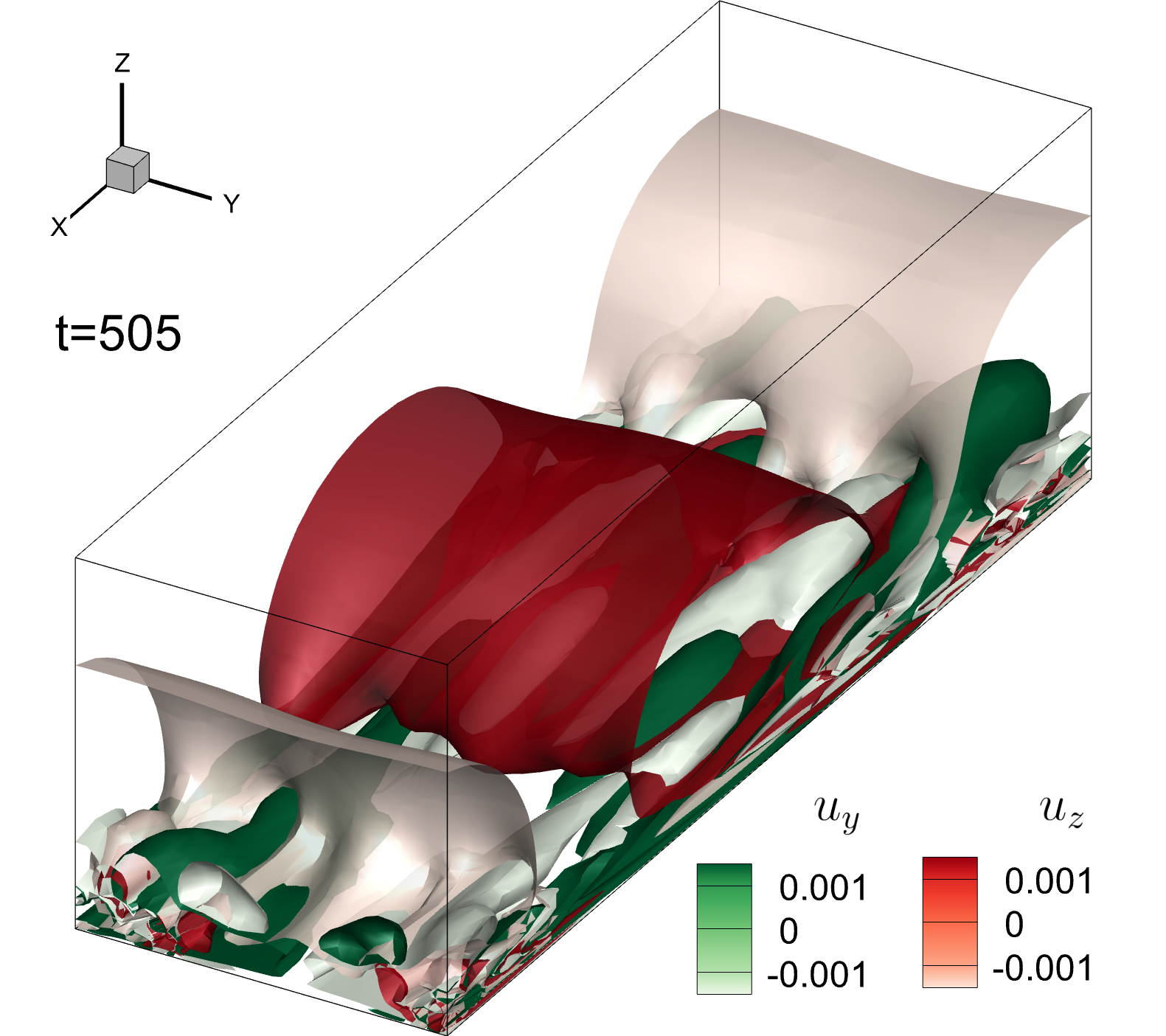}
    \includegraphics[width=0.45\linewidth]{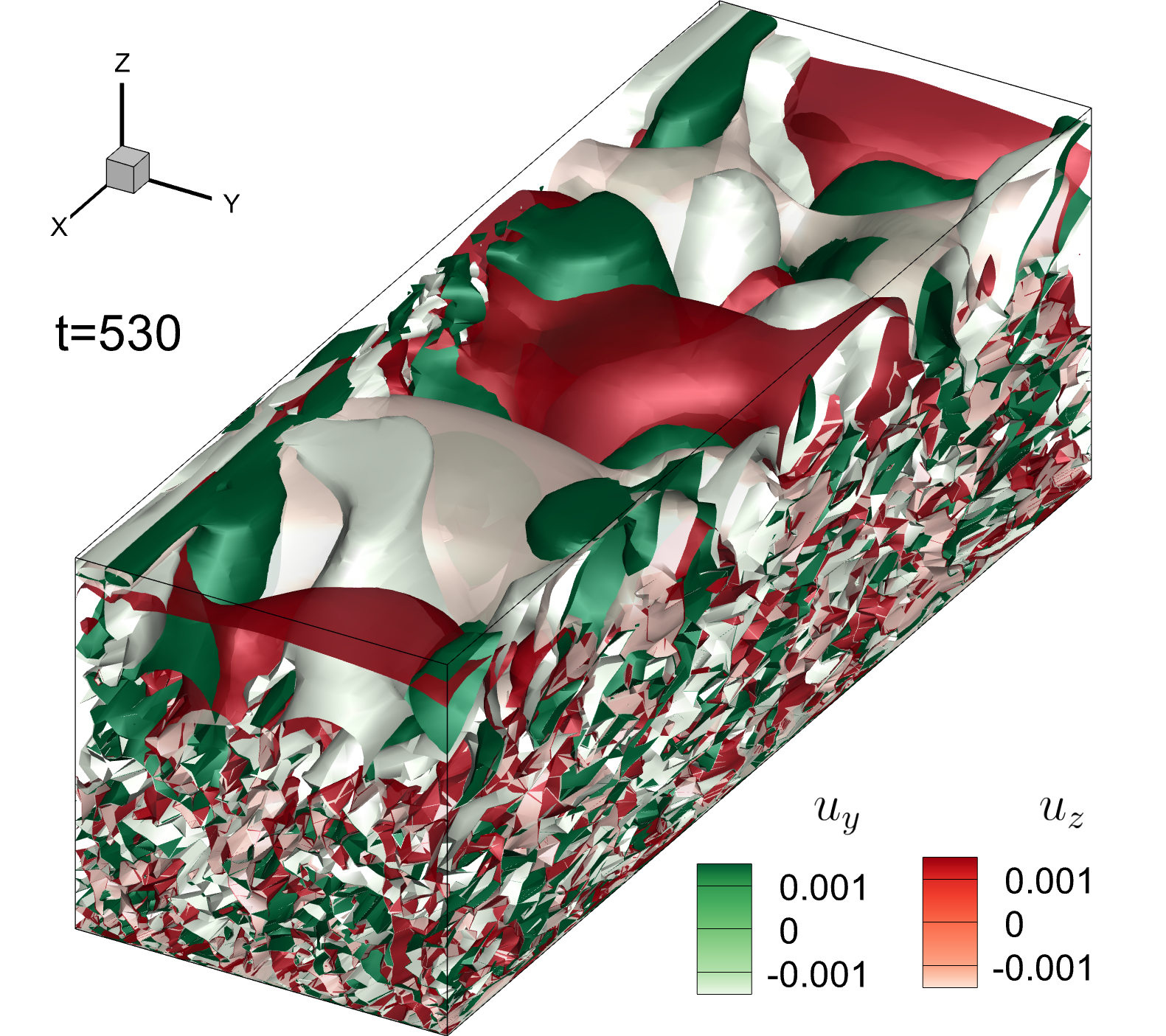}
    \caption{DNS of transition to turbulence at $\Ha=50$, $\Rey=25000$, $\varkappa=0.7$. Iso-surfaces of perturbations of $u_z$ (in red) and $u_y$ (green) at various times are shown.}
    \label{fig:frames}
\end{figure*}

The results are presented in figures \ref{fig:dns_enrg} and \ref{fig:frames}. The time history of the volume-averaged kinetic energy in each component of velocity perturbations (see figure \ref{fig:dns_enrg}) shows initial decay (up to $t\approx 100$) followed by nearly exponential growth, which is visible in $u_z$ starting at $t\approx 150$ and in $u_x$ starting at $t\approx 300$. The growth rate approaches the constant value $\gamma_i=0.013$ at $t \lesssim 450$, which differs from the result of the linear stability analysis (black dashed line in figure \ref{fig:dns_enrg}b) by less than 10\%. The isosurfaces of $u_z$ shown for $t=450$ in figure \ref{fig:frames} confirm that at this stage the fastest growing perturbations are the spanwise-independent IPI modes (compare with figure \ref{fig:eigenmodes_k07}c). 

The next stage of flow evolution is the secondary instability identifiable by the increase in the growth rate of the spanwise velocity component $u_y$ at $t\gtrsim 350$ (see figure \ref{fig:dns_enrg}b).  The secondary growing perturbations take the form of streamwise-elongated structures located near the wall $z=-1$. This indicates that this stage of transition is associated with the instability of the shear layer developing as a result of the growth of IPI modes (see figure \ref{fig:eigenmodes_k07}c). At $t\approx490$, the energy of perturbations of $u_y$ becomes comparable to the energy of perturbations of the $u_z$ component. 

The final stage of evolution is driven by nonlinear interactions and the distortion of IPI modes (see figure \ref{fig:frames} at $t=505$), along with a sharp increase in the growth rate of perturbations. The flow becomes fully turbulent at $t \approx 530$.

Although a thorough analysis of the possibility of subcritical solutions was not conducted, it is worth noting that all DNS runs performed were consistent with a supercritical bifurcation scenario: all simulations at $\Rey < \Rey_{cr}$ resulted in the decay of perturbations, while those at $\Rey > \Rey_{cr}$ exhibited linear growth followed by a transition to turbulence.

\section{Conclusions}

We have considered the influence of the inhomogeneity of the fluid properties on the Hartmann channel flow. The laminar velocity profile is found to become asymmetric if the viscosity $\nu$ or the electric conductivity $\sigma$ varies linearly across the channel. The effect of $\sigma$ is particularly strong and manifests itself even at moderate values of $\Ha$. The maximum velocity location shifts toward the wall, where the electric conductivity is lower.  It is also found that even weak variations of $\sigma$ lead to the development of inflection points in the velocity profile.

The implications of inflection points for instability and the transition to turbulence are investigated using linear stability analysis and direct numerical simulations (DNS). This study considers the case of uniform viscosity and linearly varying electrical conductivity.

A new mechanism of linear instability of the Hartmann channel flow, termed here inflection point instability (IPI), is identified. The characteristics of IPI are fundamentally different from those of the linear instability associated with Hartmann boundary layers (HLI). In particular, the fastest-growing IPI perturbations are spanwise rolls located in the core of the flow and having much smaller wavenumbers than their HLI counterparts. While the connection between IPI and the inflection points is not rigorously proven, it is strongly supported by several observations: the spatial structure of the growing perturbations, the correlation between the location of maximum kinetic energy and the inflection point, the values of the phase velocity of the IPI perturbations, and the existence of an inviscid limit for the IPI growth rate.

The threshold for the new instability, $\Rey_{cr}^\mathrm{IPI}$, decreases rapidly with increasing strength of the nonuniformity of $\sigma$, which in our study is characterized by the linear slope $\varkappa$. $\Rey_{cr}^\mathrm{IPI}$ becomes smaller than $\Rey_{cr}^\mathrm{HLI}$ at $\varkappa \sim 0.5$—significantly higher than the values of $\varkappa$ at which inflection points first appear. For flows with $\varkappa \gtrsim 0.7$, $\Rey_{cr}^\mathrm{IPI}$ is more than two orders of magnitude lower than $\Rey_{cr}^\mathrm{HLI}$.

The predictions of the linear stability analysis are supported by DNS. The latter reveals four stages of flow evolution: the decay of initial random perturbations, well-pronounced IPI instability with a growth rate close to that predicted by the linear theory, nonlinear interactions and secondary shear-layer instability, and the formation of a turbulent flow.

This paper is just the first exploration of the newly discovered physical effect. More work is warranted. The most interesting questions appear to be: \textit{(1)} the role of the nonlinear bypass transition initiated in boundary layers and its possible interaction with IPI; \textit{(2)} the true nonlinear dynamics of the flow, with conductivity and viscosity evolving as functions of a scalar field (temperature or admixture concentration) transported by the flow; and (3) the realization and implications of the new instability mechanism in practical applications.

The fact that the IPI threshold is significantly lower than the HLI threshold opens the possibility of a non-trivial scenario, which may develop in a channel with $\sigma$ and $\nu$ evolving as functions of a scalar, the linear distribution of which is a solution satisfying the boundary conditions. A non-isothermal flow in a channel with cold lower and hot upper walls is a good example. A laminar, asymmetric profile illustrated in Fig. \ref{fig:ubase} would form in such a flow. The profile would first lose stability due to IPI, causing the flow to become turbulent. The turbulent flow would then mix the fluid and homogenize the conductivity across the layer, leading to a transition to the Hartmann flow, the laminar solution of which is stable at the same values of $\Ha$ and $\Rey$. This would result in re-laminarization of the flow and restoration of the linear profile of $\sigma$ and the asymmetric velocity profile with inflection points, thus restarting the cycle.  If confirmed, this would represent a fundamentally new mechanism for generating large-scale, quasi-periodic oscillations in MHD flows of liquid metals, adding to the list of previously identified mechanisms such as large-scale intermittency \citep{boeck:2008} and magnetoconvective oscillations.\cite{zikanov2021}

\acknowledgments{

\textbf{Funding:} {RO and PF worked in frame of a major scientific project funded by the Ministry of Science and Higher Education of the Russian Federation (Agreement No. 075-15-2024-535 dated 23 April 2024).}

\textbf{Declaration of interests:} {The authors report no conflict of interest.}

\textbf{Data availability statement:} {The data that support the findings of this study are available from the authors at a reasonable request.}

\textbf{Author ORCIDs:} {R.~Okatev, https://orcid.org/0000-0003-2741-1531; O.~Zikanov https://orcid.org/0000-0003-3844-1779
; D.~Krasnov https://orcid.org/0000-0002-8339-7749; P.~Frick, https://orcid.org/0000-0001-7156-1583}

\textbf{Author contributions:} {All authors contribute equally.}
}

\bibliography{mhd}

\end{document}